\documentclass{article}
\usepackage{epsfig}
\usepackage{cite}
\def\bom#1{{\mbox{\boldmath $#1$}}}
\def\MSbar{$\overline{{\rm MS}}$}
\def\lapprox{\lower .7ex\hbox{$\;\stackrel{\textstyle <}{\sim}\;$}}
\def\gapprox{\lower .7ex\hbox{$\;\stackrel{\textstyle >}{\sim}\;$}}

\def\e{\epsilon}

\def\S{\, {\rm S}}

\def\CA{C_A}

\def\NF{N_F}
\def\NFZ{N_{F,V}}

\def\sab{s_{12}}
\def\sac{s_{13}}
\def\sbc{s_{23}}

\def\sabc{s_{123}}

\def\S{{\cal S}}

\begin{document}
\unitlength1cm
\begin{titlepage}
\vspace*{-1cm}
\begin{flushright}
ZU-TH 05/09, IPPP/09/31
\end{flushright}                                
\vskip 2.5cm

\begin{center}
\boldmath
{\Large\bf Two-Loop QCD Helicity Amplitudes for\\[2mm] 
(2+1)-Jet Production in Deep 
Inelastic Scattering}
\unboldmath
\vskip 1.cm
{\large T.~Gehrmann}$^a$ and {\large E.W.N.~Glover}$^b$

\vskip .7cm
{\it $^a$ Institut f\"ur Theoretische Physik, Universit\"at Z\"urich,
Winterthurerstrasse 190,\\ CH-8057 Z\"urich, Switzerland}
\vskip .4cm
{\it $^b$ Institute for Particle Physics Phenomenology, University of Durham,
South Road,\\ Durham DH1 3LE, England}
\end{center}
\vskip 2.0cm

\begin{abstract}
We derive the two-loop QCD helicity amplitudes for the processes
$\ell q \to \ell qg$ ($\ell \bar q \to \ell \bar q g$) and $\ell g \to \ell q\bar q$, 
which are the partonic reactions 
yielding $(2+1)$-jet final states  in deep inelastic lepton nucleon 
scattering. The amplitudes are obtained by analytic continuation of 
the known helicity amplitudes for $e^+e^- \to q\bar q g$.
We separate the infrared divergent and finite parts of the 
amplitudes using Catani's infrared factorization formula. 
The analytic results for the finite parts of the 
amplitudes are expressed in 
 terms of  one- and two-dimensional harmonic polylogarithms. To evaluate 
these functions numerically, we list in detail the non-trivial (and kinematic region dependent)
variable transformations one needs to perform. 
\end{abstract}
\vfill

\end{titlepage}                                                                
\newpage

\renewcommand{\theequation}{\mbox{\arabic{section}.\arabic{equation}}}

\section{Introduction}
\setcounter{equation}{0}

The study of the hadronic final state in deep inelastic lepton-nucleon
scattering (DIS) allows the determination of aspects of the nucleon structure 
which are not accessible in inclusive scattering as well as testing predictions
of quantum chromodynamics (QCD)  to a high accuracy. In particular,  the
two-plus-one-jet production~\cite{LOcalc}  rate  in DIS can be used both to
probe the gluon  distribution and to enable a precise  determination of the
strong coupling constant $\alpha_s$. 

The measurement~\cite{dijethera,forwardjets}  of these jet observables at the
electron-proton collider HERA, with a much larger  kinematical range than at
earlier fixed target experiments,  allows precision studies, and the inclusion
of next-to-leading order (NLO) corrections~\cite{graudenz1,allMC} has become
mandatory. However, the present data already highlight the limitations of the
NLO  description: the error on the extraction of $\alpha_s$ from HERA  (2+1)
jet  data is dominated~\cite{ashera}  not by the statistical uncertainty on the
data, but  by the uncertainty inherent to the NLO calculation, as estimated by 
varying renormalization and factorization scales. Given that the  statistical
precision of the data will further improve once all data from  the HERA-II run
are analysed, the theoretical description requires the inclusion of the
next-to-next-to-leading order (NNLO), i.e.\ ${\cal O}(\alpha_s^3)$,
corrections. 

The dijet production cross section 
in deep inelastic scattering at HERA is very sensitive to 
the gluon distribution in the proton. However, present determinations
of parton distribution functions~\cite{nnlopdf} at NNLO 
accuracy~\cite{nnlosplit} can only include 
data sets for observables where NNLO corrections are 
known~\cite{nnlocoeff}. Consequently, the precise HERA deep inelastic 
dijet data could be used in global NNLO determinations only once the 
NNLO QCD corrections to this observable are computed.

In terms of matrix elements, the calculation of these 
corrections  requires the computation
of three contributions: the tree level $\gamma^* i \to 4$~partons
amplitudes, the one-loop corrections to the $\gamma^* i \to
3$~partons amplitudes and the two-loop
 corrections to the  $\gamma^* i \to 2$~partons amplitudes, where $i=q,g$. 
All of these amplitudes can be obtained by analytic continuation from 
the respective amplitudes contributing to $e^+e^- \to 3$~jets at NNLO:
the tree level  $\gamma^*  \to 5$~partons~\cite{tree5p}, 
the one-loop corrections to the $\gamma^* \to
4$~partons amplitudes~\cite{onel4p},  
and the two-loop corrections to the  $\gamma^* \to
3$~partons amplitudes~\cite{3jme,3jtensor}. The analytic continuation 
of the tree level and one-loop amplitudes is rather straightforward, it 
has already been carried out in the context of the NLO corrections to 
the $(3+1)$-jet rate in DIS~\cite{nagy}. The analytic continuation of  
the two-loop amplitudes is more involved and is the topic of the present paper. 

Corrections to the deep inelastic $(2+1)$-jet cross section, 
which are the phenomenologically most relevant  applications, 
require only the computation of the  helicity averaged squared matrix element.
However, the helicity amplitudes  derived here can be applied to compute
angular correlations  between the outgoing lepton direction and the orientation
of the $(2+1)$-jet system~\cite{mirkesazimuth}. They also allow the 
calculation of jet cross sections in polarized  lepton-nucleon
collisions~\cite{mirkespol} which can constrain the polarized parton
distributions in the nucleon.  

Even though the matrix elements for all NNLO partonic subprocesses
contributing  to $(2+1)$-jet final states in DIS are now available, additional 
work is still needed before quantitative predictions can be made.  The two-loop
amplitudes computed here must be combined with the one-loop corrections to
$\gamma^* i\to 3$~partons, where one of the partons becomes collinear or  soft,
as well as  tree-level processes $\gamma^* i\to 4$~partons with two soft or
collinear partons in a way that explicitly allows all of the infrared singularities to
cancel one another. This task is usually accomplished by a subtraction 
method, which exploits the  universal factorization properties of 
QCD amplitudes in infrared soft or collinear limits to construct 
real radiation subtraction terms. These subtraction terms approximate 
the full real radiation matrix elements in all unresolved limits, but are 
sufficiently simple to be integrated analytically. Consequently, they 
can be combined with the virtual loop corrections to the matrix elements 
to yield infrared finite results for suitably defined observables. 
Several general subtraction schemes have been proposed at NLO~\cite{nlosub}.

For observables  with only final state partons,
an NNLO subtraction formalism, antenna subtraction, 
has been established in~\cite{ourant}. The antenna subtraction formalism 
constructs the subtraction terms from antenna functions, which are derived 
systematically from physical matrix elements~\cite{our2j}. This formalism 
has been applied in the computation of NNLO corrections to three-jet 
production in electron-positron 
annihilation~\cite{our3j,weinzierljet,weinzierlant}
and related event shapes~\cite{ourevent,weinzierlevent}. This formalism 
can be extended to include parton showers at higher orders~\cite{antshower}, 
thereby offering a process independent matching of fixed-order calculations 
and logarithmic resummations~\cite{resumall,bechernew},
 which had to be done on a case-by-case 
basis for individual observables~\cite{gionata} up to now. For processes 
with inital-state partons, antenna subtraction has been fully worked out
only to NLO so far~\cite{hadant}, work towards an extension to 
NNLO is in progress. 

Other approaches to perform NNLO calculations  of exclusive observables with initial
state partons are  the use of sector decomposition and a subtraction method based on the 
transverse momentum structure of the final state.  The sector decomposition 
algorithm~\cite{secdec}  analytically decomposes both phase space and loop integrals into
their  Laurent expansion in dimensional regularization, and  performs a subsequent
numerical  computation of the coefficients of this expansion. Using this formalism,  NNLO
results were  obtained for Higgs  production~\cite{babishiggs}     and vector boson
production~\cite{babisdy} at hadron colliders. Both reactions  were equally computed
independently using an NNLO subtraction  formalism exploiting the specific transverse
momentum  structure of these observables~\cite{grazzinihiggs}.

This paper is structured as follows: in Section~\ref{sec:kin}, we  describe the
kinematical situation relevant to DIS-$(2+1)$-jet production.  The  two-loop helicity
amplitudes are computed by analytic continuation from the  $\gamma^* \to 3$~partons
amplitudes in Section~\ref{sec:helicity}.  Finally, Section~\ref{sec:conc} contains a
discussion of the results and conclusions.

\section{Kinematics}
\setcounter{equation}{0}
\label{sec:kin}

The partonic subprocesses contributing to the production of two hard 
jets (besides the current jet) in DIS are lepton--quark scattering 
\begin{equation}
\ell (p_5) + q(-p_2) \longrightarrow \ell (-p_6) + q (p_1) + g(p_3)\; ,
\label{eq:lqkin}
\end{equation}
(with lepton-antiquark scattering following trivially from 
charge conjugation) and lepton--gluon scattering
\begin{equation}
\ell (p_5) + g(-p_3) \longrightarrow \ell (-p_6) + q(p_1)+ \bar q (p_2) \; .
\label{eq:lgkin}
\end{equation}
The momentum of the vector boson mediating the interaction is given by
\begin{equation}
-q^{\mu} = -p_6^{\mu}-p_5^{\mu}\; .
\end{equation}

It is convenient to define the invariants
\begin{equation}
\sab = (p_1+p_2)^2\;, \qquad \sac = (p_1+p_3)^2\;, \qquad 
\sbc = (p_2+p_3)^2\;,
\end{equation}
which fulfil
\begin{equation}
q^2  =(p_1+p_2+p_3)^2 = \sab + \sac + \sbc \equiv \sabc \; ,
\end{equation}
as well as the dimensionless invariants
\begin{equation}
x = \sab/\sabc\;, \qquad y = \sac/\sabc\;, \qquad z = \sbc/\sabc\;,
\end{equation}
which satisfy $x+y+z=1$.

In $e^+ e^- \to 3$~jet ($3j$) production, $q^2$ is time-like (hence 
positive) and all the $s_{ij}$ are also positive, which implies that 
$x,y,z$ all lie in the interval $[0;1]$, with the above constraint $x+y+z=1$.
For DIS-$(2+1)j$ production, $q^2$ is space-like (hence negative) 
and for lepton--quark scattering (\ref{eq:lqkin}) the invariants fulfil 
\begin{equation}
q^2 < 0\;,\quad s_{23} > 0\;,\quad s_{12} < 0, \quad s_{13} < 0\; ,
\end{equation}
or, equivalently,
\begin{equation}
x>0\;, \quad y>0 \;, \quad z<0\; .
\end{equation}
For lepton--gluon scattering (\ref{eq:lgkin}), one finds 
\begin{equation}
q^2 < 0\;,\quad s_{12} > 0\;,\quad s_{13} < 0, \quad s_{23} < 0\; \;,
\end{equation}
respectively,
\begin{equation}
x<0\;, \quad y>0 \;, \quad z>0\; .
\end{equation}
It is worth noting that the kinematical regions for these two processes 
are further subdivided by cuts in $x=1$ and $y=1$ (lepton--quark scattering)
or $y=1$ and $z=1$ (lepton--gluon scattering), as was first pointed out 
by Graudenz in the context of the calculation of the 
NLO matrix elements~\cite{graudenz1}. As a consequence,
it is not possible to find an analytic expression for the amplitudes, 
which covers the whole kinematic region {\it and} does not contain functions 
with implicit imaginary parts. We shall see below that the most convenient 
representation is obtained by subdividing the kinematical plane for each 
process according to the cuts into four sectors. In each of these sectors,
we obtain an analytic expression for the amplitudes with all 
imaginary parts made 
explicit; joining the different regions together, one obtains
a result which is continuous, but not analytic, across the cuts.

\section{Helicity amplitudes}
\label{sec:helicity}
\setcounter{equation}{0}

The renormalized 
amplitude $|{\cal M}\rangle$ can be written as
\begin{equation}
\label{eq:Mdef}
|{\cal M}\rangle = V^\mu \S_\mu(q;g;\bar q)\; ,
\end{equation}
where $V^\mu$ represents the lepton current and $\S_\mu$ denotes 
the hadron current. With $\bar q$, we denote the incoming quark 
(momentum $-p_2$) in lepton--quark 
scattering or the outgoing antiquark (momentum $+p_2$)
in lepton--gluon scattering.
A convenient method to evaluate the helicity amplitudes is in terms
of  Weyl--van der Waerden spinors, which is described briefly in
the appendix of~\cite{3jtensor}  and in detail in \cite{WvdW,six}.

Using the Weyl--van der Waerden spinor calculus, 
we can express the lepton current $V_\mu$ for photon exchange
in terms 
of the helicities of the incoming $l^- (p_6)$ and outgoing 
$l^-(-p_5)$.
Explicitly, 
\begin{equation}
V_\mu^\gamma(l^-(-p_5)-,l^-(p_6)-) = e\sigma_\mu^{\dot AB}p_{6\dot 
A}p_{5B}\frac{L^\gamma_{ll}}{p_4^2}\; , \qquad
V_\mu^\gamma(l^-(-p_5)+,l^-(p_6)+) = e\sigma_\mu^{\dot AB}p_{5\dot 
A}p_{6B}\frac{R^\gamma_{ll}}{p_4^2}\; ,
\end{equation}
where for photon exchange,
\begin{equation}
L^\gamma_{f_1f_2}
=R^\gamma_{f_1f_2}
=-e_{f_1} \delta_{f_1f_2}.
\end{equation}
It is straightforward to account for
charged and neutral current exchange by appropriate replacement of the 
couplings, keeping in mind that in these cases left- and right-handed 
couplings are no longer identical. The lepton current for incident 
anti-leptons follows from charge conjugation. 

The hadronic current $\S_{\mu}$ is related to the fixed helicity 
currents, $\S_{{\dot A}B}$, by
\begin{equation}
\S_{\mu}(q+;g\lambda;\overline{q}-) = R^\gamma_{f_1f_2}
 \sqrt2\, \sigma_{\mu}^{{\dot A}B} \S_{{\dot A}B}(q+;g\lambda;\overline{q}-) ,
\end{equation}
\begin{equation}
\S_{\mu}(q-;g\lambda;\overline{q}+) = L^\gamma_{f_1f_2}
 \sqrt2\, \sigma_{\mu}^{{\dot A}B} \S_{{\dot A}B}(q-;g\lambda;\overline{q}+) .
\end{equation}
As above, the gauge boson coupling is extracted 
from $\S_{{\dot A}B}$, and charged and neutral current reactions follow 
from appropriate substitutions.

The currents with the quark helicities flipped
follow from parity conservation:
\begin{equation}
\S_{{\dot A}B}(q-;g\lambda;\overline{q}+) =
( \S_{{\dot B}A}(q+;g(-\lambda);\overline{q}-))^*\ .
\end{equation}
Charge conjugation implies the following relations between currents with
different helicities:
\begin{equation}
\S_{{\dot 
A}B}(q\lambda_{q};g\lambda;\overline{q}\lambda_{\overline{q}}) =
(-1)
\S_{{\dot A}B}(\overline{q}\lambda_{\overline{q}};g\lambda;q\lambda_{q}).
\end{equation}
All helicity amplitudes are therefore related to the amplitudes with 
$\lambda_{q} = +$ and $\lambda_{\bar q} = -$.
Explicitly, we find
\begin{eqnarray}
{\S}_{\dot AB}(q+;g+;\overline{q}-)
&=&
\alpha(x,y,z)~
\frac{p_{1\dot AD}p_2^D p_{2B}}{\langle p_1p_3 \rangle\langle p_3{p}_2 \rangle}
+\beta(x,y,z)~
\frac{p_{3\dot AD} p_2^D p_{2B}}{\langle p_1p_3 \rangle\langle p_3{p}_2 \rangle}
+\gamma(x,y,z)~
\frac{p_{1\dot CB}p_3^{\dot C}p_{3\dot A}}{\langle p_1p_3 \rangle\langle p_3{p}_2 \rangle^*} \nonumber \\
&& +~\delta(x,y,z)~\frac{\langle p_1p_3\rangle^*}{\langle p_1p_3 \rangle\langle p_1{p}_2 \rangle^*}
\left(p_{1\dot AB}+p_{2\dot AB}+p_{3\dot AB}\right)
\;. 
\label{eq:helamp}
\end{eqnarray}
The other helicity amplitudes are obtained from $\S_{\dot 
AB}(q+;g+;\bar q-)$ by the above parity and charge conjugation relations.
Current conservation implies the following relation between the four helicity
coefficients,
\begin{equation}
\label{eq:deltadef}
\alpha(x,y,z)-\beta(x,y,z)-\gamma(x,y,z)-\frac{2\sabc}{\sab}~\delta(x,y,z) = 0.
\end{equation}
Equation~(\ref{eq:deltadef}) suffices to fix $\delta(x,y,z)$.

The three remaining helicity amplitude coefficients $\alpha$,
$\beta$ and $\gamma$ are vectors in colour space and
have perturbative expansions:
\begin{equation}
\Omega =  \sqrt{4\pi\alpha_{\rm em}} \sqrt{4\pi\alpha_s} \; \bom{T}^a_{ij}\, \left[
\Omega^{(0)}  
+ \left(\frac{\alpha_s}{2\pi}\right) \Omega^{(1)}  
+ \left(\frac{\alpha_s}{2\pi}\right)^2 \Omega^{(2)} 
+ {\cal O}(\alpha_s^3) \right] \;,\nonumber \\
\end{equation}
for $\Omega = \alpha,\beta,\gamma$ where the dependence on $(x,y,z)$ is  
implicit.

The helicity coefficients were extracted from a Feynman diagram calculation 
in dimensional regularization~\cite{dreg,hv}, with $d=4-2\e$
space-time dimensions, 
by applying suitable projectors to the amplitudes at the required order in 
perturbation theory. 
 Once these projectors  are applied to expose the helicity
structure, one is left with the task of computing a large number of 
two-loop integrals. Using integration-by-parts~\cite{hv,chet} 
and Lorentz-invariance~\cite{gr} identities, these were 
reduced~\cite{laporta} (making extensive use of MAPLE~\cite{maple} and
FORM~\cite{form}) to 
a small number of so-called master integrals, which were derived 
in~\cite{mi}. Insertion of these master integrals 
into the contraction of Feynman amplitudes with the projectors then yields the 
unrenormalized helicity amplitudes~\cite{3jtensor}.
A partial check 
 was made in~\cite{muwnew}, where two of the 
seven colour factors contributing to the 
$\gamma^* \to 3$~partons two-loop helicity amplitudes were derived 
by direct evaluation of the integrals (i.e.\ not using a reduction to 
master integrals), yielding full agreement with~\cite{3jtensor}.

The $3j$ amplitudes obtained in~\cite{3jtensor}
are expressed in terms of  two-dimensional
harmonic polylogarithms (2dHPLs).  The 2dHPLs are an extension of the harmonic 
polylogarithms (HPLs) of~\cite{hpl}.  Numerical
routines providing an evaluation of   the 
HPLs~\cite{grnum1,maitre} and
2dHPLs~\cite{grnum2} are available. 
The implementations apply 
to HPLs of arbitrary real~\cite{hpl,grnum1} or complex~\cite{maitre} 
arguments, but only for a 
limited range of arguments of the 2dHPLs. This 
range of arguments corresponds precisely to  
the $3j$ kinematics. Formulae for the analytic continuation 
of HPLs and 2dHPLs to ranges of argument relevant to 
 all $2\to 2$ scattering kinematics were derived 
in~\cite{ancont}. 
Given the restricted range of arguments for the 
numerical implementation of the 2dHPLs~\cite{grnum2}, 
in~\cite{ancont}  transformations to map all 
$2\to 2$ scattering kinematics into this range of arguments were derived.
These analytical continuations were 
subsequently confirmed by direct numerical evaluation~\cite{daleo}
 of the relevant master integrals~\cite{mi} in the relevant kinematical
regions using the numerical Mellin-Barnes technique.
Using the results of~\cite{ancont}, we can perform the analytic continuation 
of all the unrenormalized helicity amplitudes from 
the $\gamma^* \to 3$~partons kinematics considered in~\cite{3jme} to 
the kinematical situation of DIS-$(2+1)$-jet production.

Subsequently, renormalization is performed in the $\overline{{\rm MS}}$-scheme
by replacing 
the bare coupling $\alpha_0$ with the renormalized coupling 
$\alpha_s\equiv \alpha_s(\mu^2)$,
evaluated at the renormalization scale $\mu^2$
\begin{equation}
\alpha_0\mu_0^{2\e} S_\e = \alpha_s \mu^{2\e}\left[
1- \frac{\beta_0}{\e}\left(\frac{\alpha_s}{2\pi}\right) 
+\left(\frac{\beta_0^2}{\e^2}-\frac{\beta_1}{2\e}\right)
\left(\frac{\alpha_s}{2\pi}\right)^2+{\cal O}(\alpha_s^3) \right]\; ,
\end{equation}
where
\begin{displaymath}
S_\e =(4\pi)^\e e^{-\e\gamma}\qquad \mbox{with Euler constant }
\gamma = 0.5772\ldots
\end{displaymath}
and $\mu_0^2$ is the mass parameter introduced 
in dimensional regularization~\cite{dreg,hv} to maintain a 
dimensionless coupling 
in the bare QCD Lagrangian density; $\beta_0$ and $\beta_1$ are the first 
two coefficients of the QCD $\beta$-function:
\begin{equation}
\beta_0 = \frac{11 \CA - 4 T_R \NF}{6},  \qquad 
\beta_1 = \frac{17 \CA^2 - 10 C_A T_R \NF- 6C_F T_R \NF}{6}\;,
\end{equation}
with the QCD colour factors
\begin{equation}
\CA = N,\qquad C_F = \frac{N^2-1}{2N},
\qquad T_R = \frac{1}{2}\; .
\end{equation}
For the remainder of this paper we will set the renormalization scale
$\mu^2 = -q^2$. 

This procedure yields the renormalized tensor coefficients,
\begin{eqnarray}
\Omega^{(0)}  &=& \Omega^{(0),{\rm un}} ,
 \nonumber \\
\Omega^{(1)}  &=& 
S_\e^{-1} \Omega^{(1),{\rm un}} 
-\frac{\beta_0}{2\e} \Omega^{(0),{\rm un}}  ,  \nonumber \\
\Omega^{(2)} &=& 
S_\e^{-2} \Omega^{(2),{\rm un}}  
-\frac{3\beta_0}{2\e} S_\e^{-1}
\Omega^{(1),{\rm un}}  
-\left(\frac{\beta_1}{4\e}-\frac{3\beta_0^2}{8\e^2}\right)
\Omega^{(0),{\rm un}}\;.
\end{eqnarray}
The full scale dependence of the renormalized
helicity coefficients can be recovered from 
the renormalization group equation:
\begin{eqnarray}
\Omega &=& \sqrt{4\pi\alpha_{{\rm  em}}} \sqrt{4\pi\alpha_s} \; {\bom
T}^{a}_{ij}\, \bigg\{
\Omega^{(0)}  + \left(\frac{\alpha_s(\mu^2)}{2\pi}\right)
\left[
\Omega^{(1)}  
+\frac{\beta_0}{2} \Omega^{(0)} \ln\left({\mu^2\over q^2}\right)
\right]
\nonumber \\
&&+ \left(\frac{\alpha_s(\mu^2)}{2\pi}\right)^2
\bigg[\Omega^{(2)} 
+\biggl(\frac{3\beta_0}{2}\Omega^{(1)} +\frac{\beta_1}{2}\Omega^{(0)} \biggr) 
\ln\left({\mu^2\over q^2}\right)  
+\frac{3 \beta_0^2}{8} \Omega^{(0)}  
\ln^2\left({\mu^2\over q^2}\right)\bigg]
 + {\cal O}(\alpha_s^3) \bigg\}.
\end{eqnarray}

After performing ultraviolet renormalization,
the amplitudes still
contain singularities, which are of infrared origin and will be  analytically
cancelled by those occurring in radiative processes of the
same order.
Catani~\cite{catani} has shown how to organize the 
infrared pole structure of the one- and two-loop contributions renormalized in the 
\MSbar\ scheme in terms of the tree and renormalized one-loop amplitudes.
The same procedure applies to the tensor coefficients:
\begin{eqnarray}
\Omega^{(1)} &=& {\bom I}^{(1)}(\epsilon) \Omega^{(0)} +
\Omega^{(1),{\rm finite}},\nonumber \\
\Omega^{(2)} &=& \Biggl (-\frac{1}{2}  {\bom I}^{(1)}(\epsilon) {\bom I}^{(1)}(\epsilon)
-\frac{\beta_0}{\epsilon} {\bom I}^{(1)}(\epsilon) 
+e^{-\epsilon \gamma } \frac{ \Gamma(1-2\epsilon)}{\Gamma(1-\epsilon)} 
\left(\frac{\beta_0}{\epsilon} + K\right)
{\bom I}^{(1)}(2\epsilon) + {\bom H}^{(2)}(\epsilon) 
\Biggr )\Omega^{(0)}\nonumber \\
&& + {\bom I}^{(1)}(\epsilon) \Omega^{(1)}+ \Omega^{(2),{\rm finite}},
\label{eq:polesa}
\end{eqnarray}
where the constant $K$ is
\begin{equation}
K = \left( \frac{67}{18} - \frac{\pi^2}{6} \right) \CA - 
\frac{10}{9} T_R \NF.
\end{equation}
For this particular process, there is only one colour structure present at
tree level which, in terms of the gluon colour $a$ and the quark
and antiquark
colours $i$ and $j$, is simply $\bom{T}^{a}_{ij}$. Adding higher loops does not
introduce additional colour structures, and the amplitudes are therefore
vectors in a one-dimensional space.  Similarly, 
the infrared singularity operator $\bom{I}^{(1)}(\epsilon)$ is a 
scalar in  colour space
and is given by
\begin{equation}
\bom{I}^{(1)}(\epsilon)
=
- \frac{e^{\epsilon\gamma}}{2\Gamma(1-\epsilon)} \Biggl[
N \left(\frac{1}{\epsilon^2}+\frac{3}{4\epsilon}+\frac{\beta_0}{2N\epsilon}\right) 
\left({\tt S}_{13}+{\tt S}_{23}\right)-\frac{1}{N}
\left(\frac{1}{\epsilon^2}+\frac{3}{2\epsilon}\right)
{\tt S}_{12}\Biggr ]\; ,\label{eq:I1}
\end{equation}
where (since we have set $\mu^2 = -s_{123}$)
\begin{equation}
{\tt S}_{ij} = \left(\frac{s_{123}}{s_{ij}}\right)^{\epsilon}.
\end{equation}
Using the invariants $s_{ij}$ as defined in Section~\ref{sec:kin}, 
phase factors distinguishing initial and final state partons become 
obsolete in this expression. 
Note that on expanding ${\tt S}_{ij}$,
imaginary parts are generated, the sign of which is fixed by the small imaginary
part $+i0$ of all $s_{ij}$.

Finally, the term of Eq.~(\ref{eq:polesa}) that involves 
${\bom H}^{(2)}(\epsilon)$ 
produces only a single pole in $\epsilon$ and is given by 
\begin{equation}
\label{eq:htwo}
{\bom H}^{(2)}(\epsilon)
=\frac{e^{\epsilon \gamma}}{4\,\epsilon\,\Gamma(1-\epsilon)} H^{(2)} \;,  
\end{equation}
where the constant $H^{(2)}$ is renormalization-scheme-dependent.
As with the single-pole parts of $\bom{I}^{(1)}(\epsilon)$,
the process-dependent
$H^{(2)}$ can be constructed by counting the number of
radiating partons present in the event.
In our case,  in the \MSbar\ scheme:
\begin{eqnarray}
\label{eq:Htwo}
H^{(2)} &=&  
\left(4\zeta_3+\frac{589}{432}- \frac{11\pi^2}{72}\right)N^2
+\left(-\frac{1}{2}\zeta_3-\frac{41}{54}-\frac{\pi^2}{48} \right)
+\left(-3\zeta_3 -\frac{3}{16} + \frac{\pi^2}{4}\right) \frac{1}{N^2}\nonumber \\
&&
+\left(-\frac{19}{18}+\frac{\pi^2}{36} \right) N\NF 
+\left(-\frac{1}{54}-\frac{\pi^2}{24}\right) \frac{\NF}{N}+ \frac{5}{27} \NF^2.
\end{eqnarray}

At leading order 
\begin{equation}
\alpha^{(0)}(x,y,z) = \beta^{(0)}(x,y,z) = 1\qquad
\mbox{and}\qquad \gamma^{(0)}(x,y,z) =0.
\end{equation}
The renormalized next-to-leading order helicity amplitude coefficients can be 
straightforwardly obtained to all orders in $\epsilon$ from 
the known exact expressions for the one-loop master integrals. 
For practical purposes, they are needed through to ${\cal O}(\epsilon^2)$
in evaluating the 
infrared-divergent one-loop contribution to the two-loop amplitude,
while only the 
finite piece is needed for the one-loop self-interference.
They can be decomposed 
according to their colour structure as follows:
\begin{equation}
\Omega^{(1),{\rm finite}}(x,y,z) =  
N\, a_{\Omega}(x,y,z) + \frac{1}{N}\, b_{\Omega}(x,y,z) + \beta_0\, 
c_{\Omega}(x,y,z)
   \;. 
\end{equation}
The expansion of the coefficients 
through to $\e^2$ yields HPLs and 2dHPLs up to weight 4
for  $a_{\Omega}$, $b_{\Omega}$ and up to weight 3 for $c_{\Omega}$. In 
order to obtain expressions suitable for numerical evaluation (which 
requires the arguments of the 2dHPLs to be in a restricted range) and 
to make all imaginary parts explicit, we use the following 
decomposition~\cite{graudenz1} 
for the coefficients $l=a,b,c$ of the individual colour factors:
\begin{eqnarray}
l_{\Omega}(x,y,z) & = & \Theta(x)\Theta(1-x)\Theta(y)\Theta(1-y)\Theta(-z)
l_{\Omega,1}\left(x+z,-z\right) \nonumber \\
&& + \Theta(x-1)\Theta(y)\Theta(1-y)\Theta(-z)
l_{\Omega,2}\left(-\frac{y+z}{x},\frac{y}{x}\right)\nonumber \\
&& + \Theta(x)\Theta(1-x)\Theta(y-1)\Theta(-z)
l_{\Omega,3}\left(\frac{y+z}{y},\frac{x}{y}\right) \nonumber \\
&&+ \Theta(x-1)\Theta(y-1)\Theta(-z)
l_{\Omega,4}\left(\frac{y+z}{z},-\frac{1}{z}\right) 
\end{eqnarray}
for the lepton--quark scattering process (\ref{eq:lqkin}) and 
\begin{eqnarray}
l_{\Omega}(x,y,z) & = & \Theta(z)\Theta(1-z)\Theta(y)\Theta(1-y)\Theta(-x)
l_{\Omega,5}\left(x+z,-x\right) \nonumber \\
&& + \Theta(z-1)\Theta(y)\Theta(1-y)\Theta(-x)
l_{\Omega,6}\left(-\frac{x+y}{z},\frac{y}{z}\right)\nonumber \\
&& + \Theta(z)\Theta(1-z)\Theta(y-1)\Theta(-x)
l_{\Omega,7}\left(-\frac{x+z}{y},\frac{z}{y}\right) \nonumber \\
&&+ \Theta(z-1)\Theta(y-1)\Theta(-x)
l_{\Omega,8}\left(\frac{x+y}{x},-\frac{1}{x}\right) 
\end{eqnarray}
for the lepton--gluon scattering process (\ref{eq:lgkin}). 

%The regions $1,\ldots,8$ correspond to 
%1d,2c,3b,4d,1b,4c,3c,2d in our analytic continuation paper.

The coefficients $l_{\Omega,i}(r,s)$ are then expressed in terms of HPLs 
of argument $s$ and 2dHPLs of argument $r$, with $s$ featuring in the 
index vector. The above choices of argument ensure that $0\leq r\leq 1-s$ and 
$0\leq s \leq 1$, as required for the numerical evaluation of the 
2dHPLs~\cite{grnum2}. The 
explicit expressions are of considerable size, such that we shall not quote 
them here. 
An example of the size and structure of those coefficients
can be found in~\cite{3jme}, where we explicitly list the helicity-averaged
one-loop times one-loop and tree times two-loop matrix elements for 
$e^+e^- \to 3$~jets.
It should be noted that these finite pieces of the one-loop coefficients 
can equally well be written in terms of ordinary logarithms and dilogarithms,
see~\cite{ert1,gg}. The reason for expressing them in terms of HPLs and 2dHPLs 
here is their usage in the infrared counter-term of the two-loop coefficients, 
which cannot be fully expressed in terms of logarithmic and polylogarithmic 
functions. 

The finite two-loop remainder is obtained by subtracting the
predicted infrared structure (expanded through to ${\cal O}(\epsilon^0)$) from
the renormalized helicity coefficient.  We further decompose the 
finite remainder according to the colour structure, as follows:
\begin{eqnarray}
\Omega^{(2),{\rm finite}}(x,y,z) &=&  
N^2 A_\Omega(x,y,z) + B_\Omega(x,y,z) + \frac{1}{N^2} C_\Omega(x,y,z) 
+ N\NF D_\Omega(x,y,z) \nonumber \\ &&
+ \frac{\NF}{N} E_\Omega(x,y,z) + \NF^2 F_\Omega (x,y,z)
+ \NFZ \left(\frac{4}{N}-N\right) G_\Omega(x,y,z) \; , 
\end{eqnarray}
where the last term is generated by graphs where the virtual gauge boson does not
couple directly to the final-state quarks.   This contribution is denoted
by $\NFZ$ and is proportional
to the charge weighted  sum of the quark flavours. 
In the case of purely electromagnetic interactions we find,
\begin{equation}
N_{F,\gamma} = \frac{\sum_q e_q}{e_q}\; .
\end{equation}
Including  electroweak vector boson
interactions, the same class of diagrams yields not only a 
contribution from the  vector component (obtained by appropriate replacement 
of the couplings in the above formula~\cite{3jtensor}), but 
also a contribution involving the axial couplings of the 
vector boson under consideration. This 
contribution vanishes however if summed over isospin doublets. 

We apply the same decomposition as used at one loop  
for the coefficients $L=A,\ldots,G$ of the individual colour factors:
\begin{eqnarray}
L_{\Omega}(x,y,z) & = & \Theta(x)\Theta(1-x)\Theta(y)\Theta(1-y)\Theta(-z)
L_{\Omega,1}\left(x+z,-z\right) \nonumber \\
&& + \Theta(x-1)\Theta(y)\Theta(1-y)\Theta(-z)
L_{\Omega,2}\left(-\frac{y+z}{x},\frac{y}{x}\right)\nonumber \\
&& + \Theta(x)\Theta(1-x)\Theta(y-1)\Theta(-z)
L_{\Omega,3}\left(\frac{y+z}{y},\frac{x}{y}\right) \nonumber \\
&&+ \Theta(x-1)\Theta(y-1)\Theta(-z)
L_{\Omega,4}\left(\frac{y+z}{z},-\frac{1}{z}\right) 
\end{eqnarray}
for the lepton--quark scattering process (\ref{eq:lqkin}) and 
\begin{eqnarray}
L_{\Omega}(x,y,z) & = & \Theta(z)\Theta(1-z)\Theta(y)\Theta(1-y)\Theta(-x)
L_{\Omega,5}\left(x+z,-x\right) \nonumber \\
&& + \Theta(z-1)\Theta(y)\Theta(1-y)\Theta(-x)
L_{\Omega,6}\left(-\frac{x+y}{z},\frac{y}{z}\right)\nonumber \\
&& + \Theta(z)\Theta(1-z)\Theta(y-1)\Theta(-x)
L_{\Omega,7}\left(-\frac{x+z}{y},\frac{z}{y}\right) \nonumber \\
&&+ \Theta(z-1)\Theta(y-1)\Theta(-x)
L_{\Omega,8}\left(\frac{x+y}{x},-\frac{1}{x}\right) 
\end{eqnarray}
for the lepton--gluon scattering process (\ref{eq:lgkin}).

The helicity coefficients contain HPLs and 2dHPLs up to weight 4  in the
$A,B,C,G$-terms, up to weight 3 in the $D,E$-terms  and up to  weight 2 in the
$F$-term. The size of each helicity coefficient is
comparable  to the size of the helicity-averaged tree times two-loop matrix
element quoted in~\cite{3jme}. We therefore refrain from quoting them
explicitly.   The complete set of one-loop and two-loop 
coefficients in FORM and 
FORTRAN format is provided with the sources of the paper on 
the archive http://arxiv.org.

\section{Conclusions and Outlook}
\label{sec:conc}
\setcounter{equation}{0}

In this paper, we have applied the analytic continuation procedures  for
two-loop four-point functions with one off-shell leg derived  in~\cite{ancont}
to obtain the two-loop helicity amplitudes  relevant to DIS-(2+1)~jet
production from the amplitudes for $\gamma^*\to q\bar q g$ derived earlier. 
Besides enabling the computation of the  virtual two-loop corrections to the
DIS-(2+1)~jet cross section and  related event shape observables (which follow
by interfering the amplitudes  with their tree level counterparts and summing
over all external helicities), knowledge of the helicity amplitudes provides
additional information about the scattering process.  In particular, oriented
event shapes as well as jet rates in polarized DIS can be computed from them. 
These matrix elements represent one of the missing ingredients for 
 NNLO computations of deep inelastic jet production.
Such calculations require the
construction of a parton-level event generator program, which contains all 
partonic channels contributing to the observable under consideration. Since
 the one-loop and tree-level contributions at NNLO contain more partons 
than hard jets, they will contribute infrared real radiation singularities,
which have to be subtracted in the corresponding partonic channels, and 
added to the two-loop virtual contribution to yield finite expressions 
suitable for numerical evaluation. While a formalism for NNLO subtraction 
is well established for processes involving only final-state partons, 
some extra work is required to subtract singularities arising from 
unresolved emission off initial-state partons. Work in this direction is 
ongoing.

\section*{Acknowledgements}

We would like to thank Lee Garland, Thanos Koukoutsakis and particularly Ettore Remiddi for their collaboration on earlier work that lead towards this paper. 
EWNG gratefully acknowledges the support of the Wolfson Foundation and the Royal Society. This research was supported in part by the Swiss National Science Foundation (SNF) under contract 200020-117602, by the UK Science and Technology Facilities Council and  by the European Commission's Marie-Curie Research Training Network under contract MRTN-CT-2006-035505 ``Tools and Precision Calculations for Physics Discoveries
at Colliders''.

\end{document}